\documentclass[fdp,a4paper,fleqn%
]{w-art}
\def\beq{\begin{equation}}
\def\eeq{\end{equation}}
\def\beqa{\begin{eqnarray}}
\def\eeqa{\end{eqnarray}}
\def\n{\nonumber \\}

\def\dag{\dagger}
\newcommand{\id}{{1\!\!1}} 
\newcommand {\tr}{{\rm tr}\,}

\begin{document}
\DOIsuffix{theDOIsuffix}
\Volume{55}
\Month{01}
\Year{2007}
\pagespan{1}{}



\title[Phenomenological studies in the matrix models]{Phenomenological studies in the matrix models}


\author[Hajime Aoki]{Hajime Aoki\inst{1}%
  \footnote{\quad E-mail:~\textsf{haoki@cc.saga-u.ac.jp}
}}
\address[\inst{1}]{Department of Physics, Saga University, Saga 840-8502, Japan}
\begin{abstract}
Matrix models are a promising candidate 
for a nonperturbative formulation of the superstring theory.
It is possible to study how the standard model and other phenomenological models
appear from the matrix model, and estimate the probability distribution of their appearance.
This article mainly addresses studies in 
toroidal compactifications with magnetic fluxes. 
\end{abstract}
\maketitle                   






\section{Introduction}

The standard model (SM) of particle physics agrees well with experiments and is successful. 
When exploring phenomenological models beyond the SM, some guides may be helpful.
On the other hand, the SM is unsatisfactory as a final theory,
and the string theory is expected to be an ultimate theory including gravity.
String-inspired phenomenologies
have been studied extensively
(see, for instance, ref.~\cite{Blumenhagen:2005mu,Blumenhagen:2006ci,Ibanez:2012zz}).
However, the string theory has too many vacua, 
and in order to see which vacuum is realized,
we need a more underlying formulation.

A candidate for it is the matrix model (MM) 
\cite{Banks:1996vh,IKKT}.
Since the MM has a definite action and a measure,
we can compare the string vacua dynamically, and 
calculate everything, in principle. 
Indeed, in the IIB matrix model, spacetime structures have been analyzed intensively,
and four-dimensionality seems to be preferred \cite{Aoki:1998vn,Nishimura:2001sx,Kim:2011cr}.
It is also possible to study how the SM and phenomenological models appear from the MM,
and estimate the probability distribution of their appearance.

An important ingredient of the SM is the chirality of fermions.
We usually obtain chiral fermions on our spacetime 
by introducing a nontrivial topology in the extra dimensions.
There have been several works about obtaining 
chiral fermions and the SM matter content from the MM, by using
\begin{enumerate}
\item orbifoldings \cite{Aoki:2002jt,Chatzistavrakidis:2010xi}, which amount to
imposing nontrivial identification in the extra dimensions,
\item intersecting fuzzy spheres \cite{Chatzistavrakidis:2011gs,Steinacker:2014fja,Aoki:2014cya}, 
which are similar to the intersecting D-branes 
in the string theory context,
\item toroidal compactifications with magnetic fluxes \cite{Aoki:2010gv,Aoki:2012ei,Aoki:2013pba}, which are close to
the magnetized D-branes wrapping on a torus in the string theory context.
\end{enumerate}
The last two cases are related to each other by T-duality in the string theory.
This article mainly addresses the third one. 

\section{Topological configurations on a torus}

We begin with a brief review of the IIB MM \cite{IKKT,Aoki:1998bq}.
Its action has a simple form
\beq
S_{\rm IIBMM}  = -\frac{1}{g^2_{\rm IIBMM}}  ~\tr
\left({1\over 4}[A_{M},A_{N}][A^{M},A^{N}]
+{1\over 2}\bar{\Psi}\Gamma ^{M}[A_{M},\Psi ]\right) \ ,
\label{IIBMMaction}
\eeq
where $A_{M}$ and $\Psi$ are $N \times N$ Hermitian matrices.
They are also a ten-dimensional vector and a Majorana-Weyl spinor, respectively.
Performing a kind of functional integration
as a statistical system, and taking a suitable 
large-$N$ limit, one can obtain a
nonperturbative formulation of string theory.
Since the measure as well as the action is defined definitely,
we can calculate everything in principle.

Another notable feature is that both spacetime and matter 
emerge from the matrices.
While it is a nice feature as an ultimate theory,
a precise way of embedding spacetime and matter into matrices has not been
established completely.
In the reduced model \cite{Eguchi:1982nm}, where extended spaces are described by 
matrices on a point, eigenvalue distributions of the bosonic matrices $A_M$
are interpreted as momentum of the fields.
Curved spaces can also be described by 
interpreting the matrices as differential operators \cite{Hanada:2005vr}.
In their T-dual picture, the eigenvalue distributions are identified as spacetime coordinates \cite{Aoki:1998vn},
where the supersymmetry of the IIB MM is properly defined.
As a third possibility, instead of the eigenvalue distributions, one can consider noncommutative (NC) backgrounds with
$[A_M, A_N] \neq 0$, where NC space and matter fields on it are
described in the MM rather elegantly \cite{Connes:1997cr,Aoki:1999vr}.

We now consider toroidal compactifications of $M^4 \times T^6$
with $T^6$ carrying magnetic fluxes.
We here use a finite-unitary-matrix formulation for NC tori.
It is defined by the twisted reduced model 
\cite{Eguchi:1982nm,GonzalezArroyo:1982ub,GonzalezArroyo:1982hz}
(see, for instance, refs.~\cite{Ambjorn:1999ts,Ambjorn:2000nb,Ambjorn:2000cs}).
We then consider background configurations 
\beqa
A_\mu &\sim& x_\mu \otimes \id \ , \n
e^{i A_i} &\sim& \id \otimes V_i \ ,
\label{Areldec}
\eeqa
with $\mu=0,\ldots,3$ and $i=4,\ldots,9$,
where $A_M$ stand for the Hermitian matrices in the IIB MM (\ref{IIBMMaction}).
Unitary matrices $V_i$ represent $T^6$, while
$x_\mu$ represent our spacetime $M^4$.
One can alternatively consider situations where our spacetime is also compactified,
a ten-dimensional NC torus with an anisotropy of sizes
between four and six dimensions.
We will hereafter study the extra-dimensional space $T^6$
in the unitary MM.

We then focus on $V_i$ in (\ref{Areldec}).
It is known that
each topological sector is defined by the module in NC geometries.
In the MM formulations, defining module corresponds to 
imposing twisted boundary conditions on the matrices
\cite{Szabo:2001kg,Aoki:2008ik}.
It was shown in ref.~\cite{Aoki:2010gv}, however, that instead of imposing twisted boundary conditions by hand,
topological sectors can be defined by considering background matrix configurations
as follows:
\beqa
V_{3+j} &=& 
\begin{pmatrix}
 \Gamma_{1,j}^1 \otimes \id_{n^1_2} \otimes \id_{n^1_3} \otimes \id_{p^1}&&\cr
& \ddots& \cr
&& \Gamma_{1,j}^h \otimes \id_{n^h_2} \otimes \id_{n^h_3}\otimes \id_{p^h}
\end{pmatrix} \ , \n
V_{5+j} &=& 
\begin{pmatrix}
\id_{n^1_1} \otimes  \Gamma_{2,j}^1 \otimes \id_{n^1_3} \otimes \id_{p^1}&&\cr
& \ddots& \cr
&& \id_{n^h_1} \otimes \Gamma_{2,j}^h \otimes \id_{n^h_3}\otimes \id_{p^h}
\end{pmatrix} \ , \n
V_{7+j} &=&
\begin{pmatrix}
\id_{n^1_1} \otimes \id_{n^1_2} \otimes \Gamma_{3,j}^1  \otimes \id_{p^1}&&\cr
& \ddots& \cr
&& \id_{n^h_1} \otimes \id_{n^h_2} \otimes \Gamma_{3,j}^h \otimes \id_{p^h}
\end{pmatrix} \ ,
\label{conf_V6}
\eeqa
with $j=1,2$.
They are block-diagonal matrices.
Each block is a tensor product of four factors.
The first three factors each represent $T^2$ in $T^6=T^2 \times T^2 \times T^2$,
and the last factor provides a gauge group structure.
The configuration (\ref{conf_V6}) gives the gauge group
${\rm U}(p^1) \times {\rm U}(p^2) \times \cdots \times {\rm U}(p^h)$.
The unitary matrices $\Gamma_{l,j}^a$
represent NC $T^2$ with magnetic flux $q_{l}^a$,
where $a=1,\ldots,h$ labels the block, and $l=1,2,3$ labels $T^2$ in $T^6=T^2 \times T^2 \times T^2$.
They are defined by using the Morita equivalence.
For details, see, for instance,
refs.~\cite{Aoki:2010gv,Szabo:2001kg,Aoki:2008ik}.
We note that the configurations (\ref{conf_V6}) are classical solutions of
the unitary MM.

The fermionic matrix $\Psi$ is similarly decomposed into blocks as
\beq
\Psi=
\begin{pmatrix}
\varphi ^{11} \otimes \psi^{11}  & \cdots & \varphi ^{1h} \otimes  \psi^{1h} \cr
\vdots  & \ddots & \vdots \cr
\varphi ^{h1} \otimes \psi^{h1} & \cdots & \varphi ^{hh} \otimes \psi^{hh}
\end{pmatrix} \ ,
\label{psi_block_decompose}
\eeq
where $\varphi^{ab}$ and  $\psi^{ab}$ represent spinor fields on $M^4$
and $T^6$, respectively.
Each block $\varphi^{ab} \otimes \psi^{ab}$ is in a bifundamental representation
$(p^a,\bar{p^b})$ under the gauge group ${\rm U}(p^a) \times {\rm U}(p^b)$.
It turns out \cite{Aoki:2010gv}
that $\psi^{ab}$ has a topological charge in $T^6$ as 
\beq
p^a p^b \prod_{l=1}^3 (q^a_l-q^b_l)  \ .
\label{indexab}
\eeq
Indeed,
by using an overlap-Dirac operator,
which satisfies a Ginsparg-Wilson relation and an index theorem,
the Dirac index, i.e., 
the difference between the numbers of chiral zero modes,
was shown to take the values (\ref{indexab}).  

\section{Phenomenological studies}

Let us start phenomenological studies.
We first find all the matrix configurations (\ref{conf_V6}) that provide 
phenomenological models where
\begin{itemize}
\item[1.] fermion matter content is exactly the SM one plus a right-handed neutrino with 
replication of three generations,
\item[2.] gauge group includes the SM one, ${\rm SU}_c(3) \times {\rm SU}_L(2) \times {\rm U}_Y(1)$, and  is a subgroup of U(8). 
\end{itemize}
It turns out \cite{Aoki:2013pba} that the following four gauge groups exhaust all the possibilities:
\begin{itemize}
\item[(i)] ${\rm U}(4) \times {\rm U}_L(2) \times {\rm U}_R(2)$
\item[(ii)] ${\rm U}_c(3) \times {\rm U}_l(1) \times {\rm U}_L(2) \times {\rm U}_R(2)$
\item[(iii)]${\rm U}(4) \times {\rm U}_L(2) \times {\rm U}(1)^2$
\item[(iv)]${\rm U}_c(3) \times {\rm U}_l(1) \times {\rm U}_L(2) \times {\rm U}(1)^2$
\end{itemize}
In the last case, ${\rm U}_c(3) \simeq {\rm SU}_c(3) \times {\rm U}(1)$ gives the color ${\rm SU}_c(3)$,
and $ {\rm U}_L(2)  \simeq {\rm SU}_L(2) \times {\rm U}(1)$ gives ${\rm SU}_L(2)$ of the electroweak interaction.
${\rm U}_l(1)$ corresponds to the lepton number.
There are five ${\rm U}(1)$'s in total, whose linear combinations give the hypercharge  ${\rm U}_Y(1)$
and extra ${\rm U}(1)$'s.
In case (iii), the color ${\rm U}_c(3)$ and the lepton number ${\rm U}_l(1)$ are unified to ${\rm U}(4)$,
which reminds us of the Pati-Salam model \cite{Pati:1974yy}.
In case (ii), the ${\rm U}(1)^2$ are unified to ${\rm U}_R(2)$.
Since it acts on the right-handed fermions, we denote it with the subscript $R$.
In case (i), both ${\rm U}(4)$ and ${\rm U}_R(2)$ unifications take place.
There are several solutions in each case.
Explicit forms of the solutions are given in ref.~\cite{Aoki:2013pba}.

We next estimate probability distribution of their appearance,
by performing semiclassical analyses in the MM.
We consider the unitary MM with the bosonic action
\beq
S_{b} = -\beta {\cal N} \, \sum_{i \ne j} 
{\cal  Z}_{ji}
~\tr ~\Bigl({\cal V}_i\,{\cal V}_j\,{\cal V}_i^\dag\,{\cal V}_j^\dag\Bigr)  \ ,
\label{unitaryMM}
\eeq
where the unitary matrices ${\cal V}_i$ correspond to $\id \otimes V_i$ in (\ref{Areldec}), 
and  ${\cal N}$ is size of the matrices ${\cal V}_i$.
$\beta$ is the coupling constant, and ${\cal  Z}_{ji}$ are the twist parameters.
The unitary MM (\ref{unitaryMM}) has the same symmetries as the IIB MM (\ref{IIBMMaction}):
the Lorentz symmetry, though it is slightly broken by the noncommutativity, and U(${\cal N}$) gauge symmetry.
They both are reduced models of the Yang-Mills theory.
Then, the unitary MM can be regarded as a low-energy-effective theory of the IIB MM.
One may also use it as another definition for the string theory,
though fine tunings may be needed when taking the large-${\cal N}$ limit,
since the supersymmetry is difficult to write down in the unitary MM.
By inserting the topological configurations (\ref{conf_V6}) into (\ref{unitaryMM}),
one obtains the classical action \cite{Aoki:2012ei,Aoki:2013pba}
\beq
\Delta S_b \simeq 4 \pi^2  \beta \frac{{\cal N}^2}{k N^4}  
\sum_{l=1}^3 \sum_{a=1}^h p^a (q^a_l)^2 \ ,
\label{deltaS}
\eeq
where the difference from the minimum value is given.
${\cal N}$ and $N$ are size of the matrices ${\cal V}_i$ and $V_i$, respcetively.
$k$ represents maximal gauge group ${\rm U}(k)$, and $k=8$ in the present case.
We call (\ref{deltaS}) an instanton action
since it is a classical action of a topological configuration.

We calculate instanton actions for the matrix configurations that provide phenomenological models.
The results are given in Tables 12 and 13 in ref.~\cite{Aoki:2013pba}.
Rather small instanton actions are obtained for the phenomenological models.
Substantial differences among the instatnton actions of the phenomenological modes are not found.

\section{Conclusions and discussions}

We have studied phenomenologies in the MM's,
which are expected to be a nonperturbative formulation of the string theory.
In particular, by considering situations with the toroidal compactifications with magnetic fluxes,
we found all the matrix configurations that provide phenomenological models,
and estimated the probabilities of their appearance.

Higgs field can also be embedded in the matrices, since
the gauge fields in the extra dimensions $V_i$ give scalar fields on our spacetime.
Some blocks have the same representation under the gauge group
as the SM Higgs field.
However, it is difficult to keep scalar fields massless against quantum corrections,
which is known as the naturalness, or the hierarchy problem.
If we rely on the supersymmetry, we need to study how to realize it and how to break it 
in the MM's.
Ideas of
the large extra dimensions \cite{ArkaniHamed:1998rs} 
and the gauge-higgs unifications \cite{Manton:1979kb,Fairlie:1979at,Hosotani:1983xw,Hatanaka:1998yp}
can also be applied to these MM's. 

It is important to study how the electroweak symmetry breaking occurs in the MM's.
Ideas in the gauge-higgs unifications and the recombination of the intersecting D-branes \cite{Cremades:2002cs}
can be used.
We can also study values of the Yukawa couplings and the flavor structures
as was done in ref.~\cite{Cremades:2004wa,Abe:2012fj}.

Since the extra U(1) gauge groups arise in these phenomenological models,
there remain anomalies within the gauge dynamics.
In string theory context, the anomaly is canceled via the Green-Schwarz mechanism,
by the exchanges of the RR fields,
which also make some U(1) gauge fields massive.
It is important to study how these mechanisms are realized in the MM's.

Finally, we emphasize once again the importance of these phenomenological studies in the MM's.
First, such studies may give us a guide for exploring phenomenologies beyond the SM.
They may also give us a criterion for justifying or modifying the formulation of MM's.
Secondly, since the MM has the definite action and measure,
we can survey the probability distribution over the whole of the string vacua.

\end{document}